\documentclass[sigconf]{acmart}
\renewcommand{\abstractname}{\uppercase{Abstract}}
\usepackage{bm}
\usepackage{balance}
\usepackage{multicol} 
\usepackage{multirow}
\usepackage{booktabs}
\usepackage{enumitem} 
\usepackage{graphicx}
\usepackage{hyperref}
\usepackage{pifont}
\usepackage{tabularx}
\usepackage[table]{xcolor} 
\allowdisplaybreaks

\AtBeginDocument{%
  }

\copyrightyear{2026}
\acmYear{2026}
\setcopyright{cc}
\setcctype{by-nc-nd}
\acmConference[SIGIR '26]{Proceedings of the 49th International ACM SIGIR Conference on Research and Development in Information Retrieval}{July 20--24, 2026}{Melbourne, VIC, Australia}
\acmBooktitle{Proceedings of the 49th International ACM SIGIR Conference on Research and Development in Information Retrieval (SIGIR '26), July 20--24, 2026, Melbourne, VIC, Australia}
\acmDOI{10.1145/3805712.3809538}
\acmISBN{979-8-4007-2599-9/2026/07}

\begin{document}
\title{CARD: Non-Uniform Quantization of Visual Semantic Unit for Generative Recommendation}

\author{Yibiao Wei}
\affiliation{%
  \institution{University of Electronic Science and Technology of China}
  \city{Chengdu}
  \state{Sichuan}
  \country{China}
}
\email{weiyibiao12138@gmail.com}

\author{Jie Zou}
\authornote{Corresponding author.}
\affiliation{%
  \institution{University of Electronic Science and Technology of China}
  \city{Chengdu}
  \state{Sichuan}
  \country{China}
}
\email{jie.zou@uestc.edu.cn}

\author{Pengfei Zhang}
\affiliation{%
  \institution{University of Electronic Science and Technology of China}
  \city{Chengdu}
  \state{Sichuan}
  \country{China}
}
\email{202421080437@std.uestc.edu.cn}

\author{Xiao Ao}
\affiliation{%
  \institution{University of Electronic Science and Technology of China}
  \city{Chengdu}
  \state{Sichuan}
  \country{China}
}
\email{202422080439@std.uestc.edu.cn}

\author{Weikang Guo}
\affiliation{%
  \institution{Southwestern University of Finance and Economics}
  \city{Chengdu}
  \state{Sichuan}
  \country{China}}
\email{guowk@swufe.edu.cn}

\author{Zeyu Ma}
\affiliation{%
  \institution{University of Electronic Science and Technology of China}
  \city{Chengdu}
  \state{Sichuan}
  \country{China}}
\email{cnzeyuma@163.com}

\author{Yang Yang}
\affiliation{%
  \institution{University of Electronic Science and Technology of China}
  \city{Chengdu}
  \state{Sichuan}
  \country{China}}
\email{yang.yang@uestc.edu.cn}

\renewcommand{\shortauthors}{Yibiao Wei et al.}

\begin{abstract}
Generative recommendation frameworks typically represent items as discrete Semantic IDs (SIDs). While existing studies have sought to enhance SID construction by incorporating multimodal content, collaborative signals, or more advanced quantization techniques, learning high-quality SIDs still faces two key challenges: 
(1) The two-stage generative recommendation paradigm (SID construction and autoregressive generation) provides insufficient supervision for heterogeneous fusion, which hinders learning high-quality SIDs, and (2) non-uniform embeddings lead to codeword imbalance and generation bias.
To address these challenges, we propose a novel generative recommendation framework, called CARD. CARD introduces a visual semantic unit that unifies textual, visual, and collaborative signals into a structured visual representation prior to encoding, enabling holistic semantic modeling and effectively alleviating the semantic gap, thereby reducing the reliance on supervision signals during SID learning. Furthermore, to deal with the highly non-uniform distribution of item semantic embeddings in recommendation scenarios, we develop a non-uniform quantization framework (NU-RQ-VAE), which incorporates a learnable and invertible non-uniform transformation into the quantization process to map skewed semantic distributions into a more balanced latent space, thereby significantly improving codebook utilization and quantization accuracy. Experiments on multiple datasets show that CARD consistently outperforms baseline methods under various settings; meanwhile, the proposed non-uniform transformation module is plug-and-play and remains robust across different quantization schemes. Code is available at \url{https://github.com/HAI-UESTC/CARD}.
\end{abstract}



\maketitle

\section{INTRODUCTION}
Recommender systems \cite{SASRec,GRU4Rec,Qrec,MCCRS,GeoCRS,TSCRKG,yaoyan,lijingzhi,DivReason} play a crucial role in modern information services by connecting users with massive information resources.
Traditional recommendation methods \cite{lightgcn,HDKCL,CooSBR,yang2026,KMDCL,DisenCRS} predominantly follow an identifier (ID)-based modeling paradigm, which suffers from limited representational expressiveness and exhibits poor generalization to unseen items, particularly under sparse interaction settings \cite{cold-start1,cold_start2}.

To alleviate these limitations, generative recommendation \cite{TIGER,OneRec,HSTU} has recently emerged as a promising approach. Unlike ID-based modeling paradigms, generative recommendation represents each item with semantic IDs (SIDs) by quantizing item textual content into sequences of discrete codes \cite{RQ-VAE,OneRec,VQ-VAE}, which improves generalization under interaction sparsity and cold-start settings. Subsequently, generative models (e.g., T5 \cite{T5}) perform autoregressive generation over SIDs, formulating next-item prediction as a sequence-to-sequence task.
Existing studies of generative recommendation enhance the construction of SIDs from multiple perspectives, including both information enrichment and quantization design. 
Some approaches incorporate multimodal features \cite{MMQ,UTGRec,MACRec} and collaborative signals \cite{LC-Rec,LETTER,EAGER} into the quantization process to learn richer semantic representations. 
Beyond incorporating multimodal and collaborative signals, recent studies further focus on the quantization mechanism itself, designing more fine-grained or efficient quantization strategies (e.g., OPQ \cite{RPG} and Multi-Head VQ-VAE \cite{LlaDA-Rec}) to improve the expressiveness and generation efficiency of SIDs. Despite these advances, constructing high-quality SIDs still faces two key challenges:

\textbf{\textit{Challenge 1: Insufficient Supervision Limits Heterogeneous Information Fusion for SID Quantization.}} 
Although existing methods \cite{LETTER,MQL4GRec,MACRec} incorporate heterogeneous information such as textual, visual, and collaborative signals, most still encode each modality with independent encoders. Mainstream generative recommendation approaches further adopt a non-end-to-end two-stage paradigm \cite{LC-Rec,TIGER, OneRec}: they first learn item SIDs independently and then keep them fixed for subsequent recommendation. Since SID construction is decoupled from the final recommendation objective, heterogeneous fusion lacks direct downstream supervision. In this setting, most existing work often enforces cross-modal consistency by explicitly aligning \cite{Align3GR,LETTER,MQL4GRec} (e.g., contrastive losses \cite{LETTER,MACRec}) embeddings or SIDs across modalities. 
However, strong alignment may ignore cross-modal differences, leading to over-homogenization that weakens unimodal discriminability, and may even distort modality-specific semantics. This issue is particularly critical for quantization, where continuous semantics must be compressed into discrete codes and any semantic conflicts or dominance can be amplified during discretization. Therefore, a natural question arises: under an insufficiently supervised setting, can we unify heterogeneous semantic representations, avoiding complex fusion mechanisms, and perform holistic semantic modeling tailored for SID construction?

\begin{figure}[tbp]
  \centering
  \begin{minipage}{.36\linewidth}
    \centering
    \includegraphics[width=\linewidth]{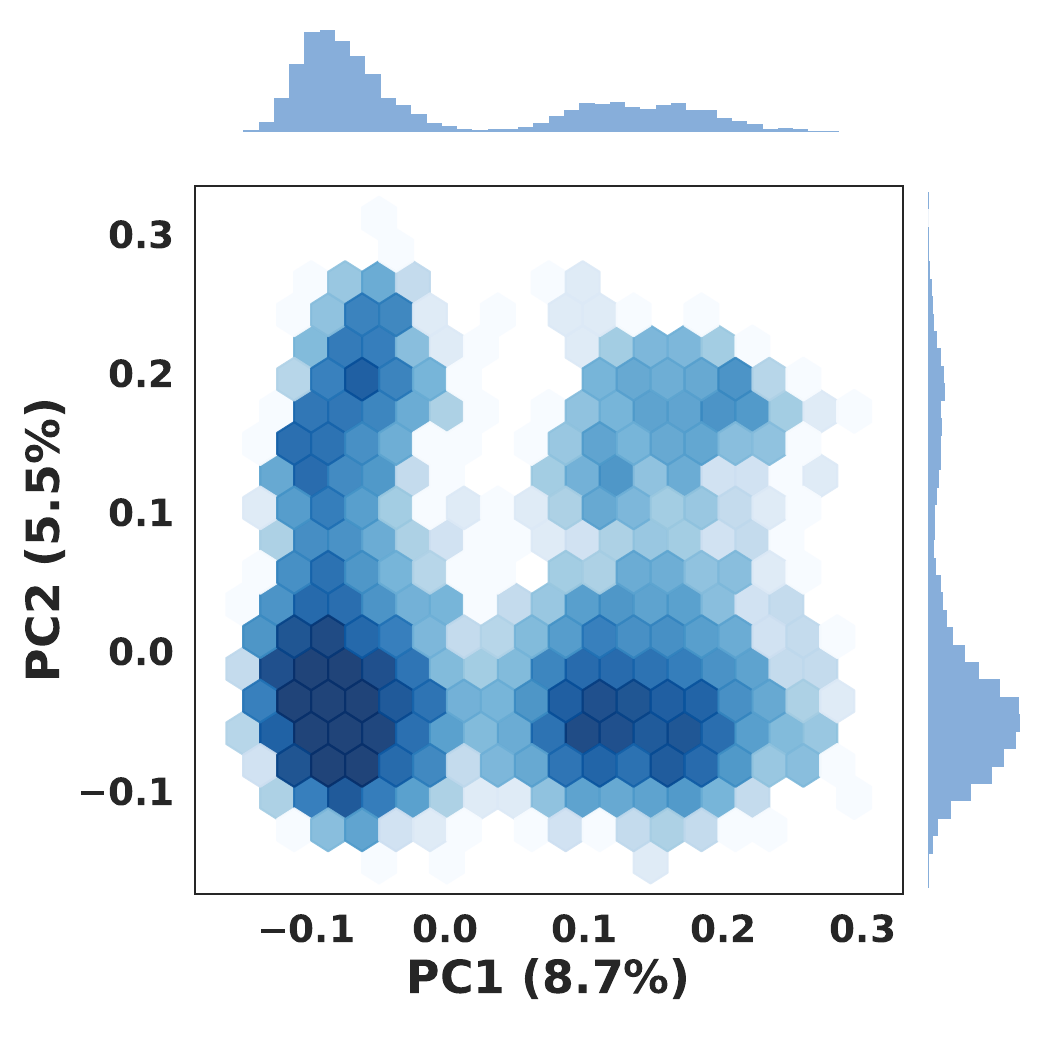}
  \end{minipage}\hfill
  \begin{minipage}{.64\linewidth}
    \centering
    \includegraphics[width=\linewidth]{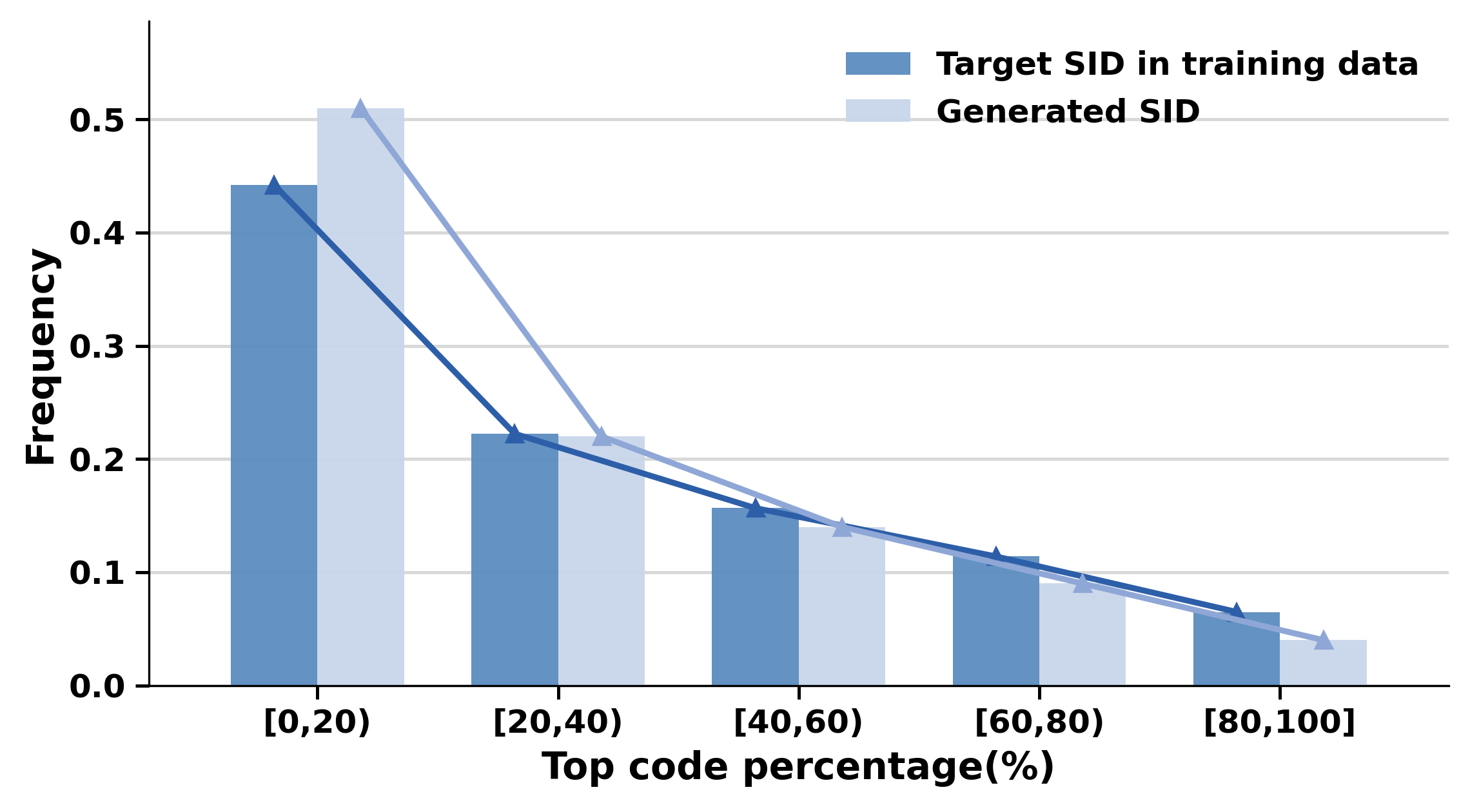}
  \end{minipage}
  \caption{The left shows item embeddings in a 2D PCA space, with dense clusters and sparse regions (darker colors indicate higher density). The right shows skewed codeword usage, where a few frequent codewords dominate and are further amplified in generated SIDs, causing generation bias.}
  \label{fig:motivation2}
\end{figure}

\textbf{\textit{Challenge 2: Non-uniform Embeddings Result in Codeword Imbalance and Generation Bias.}} 
Traditional quantizers (e.g., RQ-VAE \cite{RQ-VAE}) minimize global
reconstruction error, which is more appropriate when the embedding density is relatively uniform. 
However, Figure~\ref{fig:motivation2} (left) shows that item semantic embeddings in recommendation scenarios are highly non-uniform: semantically similar and popular items form dense clusters, while numerous long-tail items are scattered across sparse regions. 
Applying distribution-agnostic quantization to such non-uniform embedding spaces leads to imbalanced codeword assignment. 
Figure~\ref{fig:motivation2} (right) summarizes codeword usage for target and generated SIDs: a few frequent codewords dominate and are further amplified during autoregressive decoding, causing generation bias. 
From a spatial perspective, this imbalance translates into uneven effective codebook capacity across the embedding space.
In dense regions, limited codebook capacity must represent many items, hindering fine-grained discrimination, while in sparse regions, a portion of the codebook capacity remains underutilized. This imbalance ultimately limits the ability to effectively represent diverse item semantics. It naturally raises a question: can we explicitly correct the skewed embedding distribution before or during quantization to balance codeword usage and improve codebook utilization and accuracy?

To answer the above questions, we propose non-uniform quantization of visual semanti\underline{\textbf{C}} unit for gener\underline{\textbf{A}}tive \underline{\textbf{R}}ecommen\underline{\textbf{D}}ation, named \textbf{CARD}. 
The design of CARD is inspired by card-based games (e.g., \textit{Slay the Spire}\footnote{\url{https://www.megacrit.com/slay-the-spire}}), where a single card compactly integrates multiple attributes and effects into a unified representation. Motivated by this design intuition, we propose a visual semantic unit representation that organizes heterogeneous information (textual descriptions, visual content, and collaborative signals) into a structured card-style image.
By unifying heterogeneous modalities into a unified visual modality, our design enables holistic integration before encoding, effectively alleviating the semantic gap. 
As all modalities are fused within a unified visual representation space, the subsequent process avoids the need for complex fusion mechanisms, leading to improved semantic consistency and reduced system complexity. 
Although the above design enables unified semantic modeling, item embeddings in recommendation scenarios still exhibit highly skewed distributions, which poses new challenges for the subsequent quantization stage.
We further propose a Non-Uniform Residual Quantized Variational Autoencoder (NU-RQ-VAE). 
The core idea is to introduce a learnable and invertible nonlinear transformation prior to quantization, which can be viewed as an ``adjustable ruler'' that maps the original non-uniform semantic distributions into an approximately uniform latent space. 
This transformation enables subsequent residual quantization to operate under more favorable distributional conditions, thereby significantly improving codebook utilization and quantization accuracy. 
Specifically, NU-RQ-VAE provides two non-uniform transformation variants:
(1) Kumaraswamy-based parameterized transformation. This approach adopts the cumulative distribution function (CDF) of the Kumaraswamy distribution as the non-uniform transformation, enabling flexible modeling of various complex non-uniform semantic distributions during global training and achieving more thorough distribution correction.
(2) Logistic–Logit–based transformation. This approach adopts an invertible Logistic–Logit transformation tailored for bell-shaped semantic embeddings in recommendation scenarios. The main contributions of this work are summarized as follows:

\begin{itemize}[leftmargin=*]
\item We propose a novel generative recommendation method, CARD, which addresses both the insufficient supervision in heterogeneous fusion and the non-uniformity of semantic embeddings in SID construction via unified semantic modeling and non-uniform quantization.
\item We propose a visual semantic unit that unifies heterogeneous semantic signals into a visual representation, enabling holistic and quantization-friendly semantic modeling while simplifying heterogeneous information fusion.
\item We propose NU-RQ-VAE to handle non-uniform semantic embeddings in recommendation by introducing a learnable and invertible transformation into quantization, which improves codebook utilization and SID generation quality.
\item Experiments on multiple datasets show that CARD consistently outperforms existing methods, while the proposed non-uniform transformation module serves as a plug-and-play component and maintains robust performance across different quantization schemes.
\end{itemize}

\begin{figure*}[htbp]
  \centering
  \includegraphics[width=\textwidth]{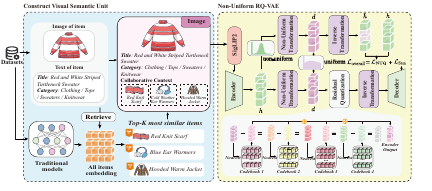}
\caption{Overview of the CARD framework. CARD constructs a visual semantic unit for each item by unifying visual, textual, and collaborative signals within the image modality. Each unit is encoded by SigLIP2 and processed by a Non-Uniform Residual Quantization Variational
Autoencoder (NU-RQ-VAE), where a non-uniform transformation and multi-level residual quantization produce discrete SIDs. These SIDs are used for downstream generative recommendation.}
\label{fig:CARD}
\end{figure*}

\section{RELATED WORK}

\subsection{Generative Recommendation}
In recent years, inspired by the success of large language models (LLMs), generative recommendation \cite{GenCDR,GRAM,GRID,TIGER,RPG,ActionPiece,GNPR-SID} has emerged as an important research direction in recommender systems. Such methods typically represent items as sequences of discrete SIDs and employ generative models (e.g., T5 \cite{T5}) to predict the SID sequence of the target item. Early studies on SID construction mainly adopted cluster-based strategies \cite{EAGER,SEATER} to discretize item embeddings. Subsequent studies introduced vector quantization techniques, including models based on residual quantization (e.g., TIGER \cite{TIGER}, LETTER \cite{LETTER}, and LC-Rec \cite{LC-Rec} based on RQ-VAE \cite{RQ-VAE}, as well as OneRec \cite{OneRec} based on RQ-KMeans) and product quantization methods \cite{RPG}, thereby enabling the construction of more expressive SIDs. Beyond quantization mechanisms, recent studies have enriched SID construction by incorporating collaborative signals \cite{LETTER,EAGER,Align3GR} and multimodal content \cite{MACRec,MMGRec,MQL4GRec,MMQ,UTGRec}. Representative studies such as EAGER \cite{EAGER} and LETTER \cite{LETTER} integrated semantic representations with collaborative signals to generate more informative and recommendation-relevant SIDs. SETRec \cite{SETRec} jointly encoded semantic and collaborative information as a set of unordered tokens and enables parallel generation, improving generation efficiency and recommendation performance. MQL4GRec \cite{MQL4GRec} mapped multimodal content into a unified quantized language, enhancing semantic representation and knowledge transfer through a shared vocabulary space and generative objectives. MMQ \cite{MMQ} and MACRec \cite{MACRec} focused on multimodal SID construction, where MMQ \cite{MMQ} modeled modality collaboration and specificity through shared–specific multi-expert quantizers, while MACRec \cite{MACRec} reduced codebook conflicts via cross-modal quantization and multi-faceted alignment. While existing methods rely on modality-specific encoders and post-hoc alignment under insufficient supervision, CARD renders heterogeneous signals into coherent visual semantic units for pre-encoding fusion and incorporates non-uniform residual quantization, enabling more stable quantization and higher-quality SID generation.

\subsection{Multi-modal Recommendation}
Multi-modal recommendation \cite{m1,m2,m3,MSCRS,FUMMER,MCCL} enhances performance by leveraging the multi-modal features of items. Earlier models \cite{VBPR,MMGCN} primarily used such signals to complement ID-based collaborative filtering. More recent studies \cite{MMGCN,MMSSL}, motivated by graph neural networks, integrated multi-modal features into graph-based learning paradigms; MMGCN \cite{MMGCN} proposed user and item representations in different modalities through specific modality graph structures.
LATTICE \cite{LATTICE} considered specific modality semantic graphs and integrated them with graph-based collaborative filtering methods. To address the semantic gap between different modalities, BM3 \cite{BM3} proposed a novel method and aligned these features through contrastive learning. Similarly, MMSSL \cite{MMSSL} employed adversarial learning to learn user and item representations across different modalities and fuse these representations through cross-modal contrastive learning. 
MENTOR \cite{Mentor} used ID-guided hierarchical alignment to bridge modality gaps while preserving interaction history. Besides contrastive learning, PromptMM \cite{PromptMM} fused user preferences from different modalities through ranking distillation \cite{RD}. MELON \cite{MELON} integrated “interaction neighbors” and “modality neighbors” to model the complex relationships between user and item modality features. Existing multimodal recommendation methods typically adopt a paradigm of modality-specific encoding followed by late-stage fusion. In non-end-to-end generative recommendation pipelines, heterogeneous signals often lack effective, task-driven fusion, which in turn poses challenges for stable SID learning and quantization. In contrast, CARD rethinks multimodal fusion and offers a new solution for SID construction.

\section{METHODOLOGY}
This section introduces the CARD framework, as illustrated in Figure~\ref{fig:CARD}. CARD first unifies heterogeneous semantic information into visual representations by constructing visual semantic units, then applies a Non-Uniform Residual Quantized Variational Autoencoder (NU-RQ-VAE) to quantize non-uniform semantic embeddings, and finally performs generative recommendation in an autoregressive manner based on the SIDs.

\subsection{Card-style Unified Item Representation}
Existing SID construction methods typically use modality-specific encoders to separately model textual, visual, and collaborative signals. This separated modeling places heterogeneous information in disjoint representation spaces, so cross-modal relations are addressed only via post-hoc fusion or alignment, creating a semantic gap during representation learning. In generative recommendation, SID learning is often insufficiently supervised and decoupled from downstream objectives; 
insufficient fusion across heterogeneous modalities can then be fixed by quantization into discrete codes, limiting SID quality.
To mitigate this problem, we propose a rendering-based fusion strategy: before encoding, we render textual attributes, visual content, and collaborative relations into a structured visual representation, allowing the encoder to process a single semantically consistent input rather than multiple independent modality channels. Next, we describe the construction of visual semantic units.

\subsubsection{\textbf{Construction of Visual Semantic Units.}}
We propose a card-style unified item representation method that integrates multimodal information and collaborative signals prior to the encoding stage. By organizing textual information, visual images, and collaborative signals into a structured visual input, the encoder can directly model holistic item semantics without requiring explicit cross-modal alignment or post-hoc fusion mechanisms. For each item $i$, we construct a card image $\mathcal{G}_i \in \mathbb{R}^{H \times W \times 3}$, which consists of three functionally complementary regions: (1) a visual region, (2) a textual semantic region, and (3) a collaborative signal region.

\textbf{Visual region.}  
The upper part of the card displays the original image of the item, denoted as $x^{\text{img}}_i$, which preserves the most intuitive and discriminative visual cues and provides foundational visual signals for subsequent semantic encoding.

\textbf{Textual semantic region.}  
The textual semantic region is designed to incorporate structured textual attributes of the item. We render the textual attributes $x^{\text{text}}_i$ (e.g., title and category) into a text panel using a fixed template and place it below the visual region to complement high-level semantic information that is difficult to capture from images alone.

\textbf{Collaborative signal region.}  
To explicitly incorporate collaborative signals into the visual semantic unit, we construct a collaborative signal region at the bottom of the card, where behaviorally similar neighbor items are presented as visual context. Specifically, we first employ a traditional recommender model (e.g., SASRec\cite{SASRec}) to learn collaborative embeddings $\mathbf{u}_i \in \mathbb{R}^d$ for each item $i \in \mathcal{I}$. For a target item $i$, we use its embedding $\mathbf{u}_i$ as the query to identify its collaborative neighborhood $\mathcal{N}(i)$. Let $s(i, j) = \mathbf{u}_i^\top \mathbf{u}_j$ be the similarity scoring function. The neighborhood $\mathcal{N}(i)$ is constructed by selecting a subset of $K_{\text{neighbors}}$ items from the item set $\mathcal{I} \setminus \{i\}$ that maximizes the sum of similarity scores. This selection process is formally defined as:
\begin{equation}
\label{eq:ann_search}
\mathcal{N}(i) = \underset{\mathcal{N} \subset \mathcal{I} \setminus \{i\}, |\mathcal{N}|=K_{\text{neighbors}}}{\operatorname{argmax}} \sum_{j \in \mathcal{N}} s(i, j),
\end{equation}
where $K_{\text{neighbors}}$ is a hyperparameter controlling the number of neighbors and we set it to 3 based on empirical validation.
The items in the resulting neighborhood set $\mathcal{N}(i)$ are arranged horizontally at the bottom of the card, with each neighbor item represented by its thumbnail image and corresponding textual title. Through this design, collaborative signals that originally reside only in the latent embedding space are transformed into visual context directly perceivable by the encoder. 

\begin{figure}[tbp]
  \centering
  \begin{minipage}{\linewidth}
    \centering
    \includegraphics[width=\linewidth]{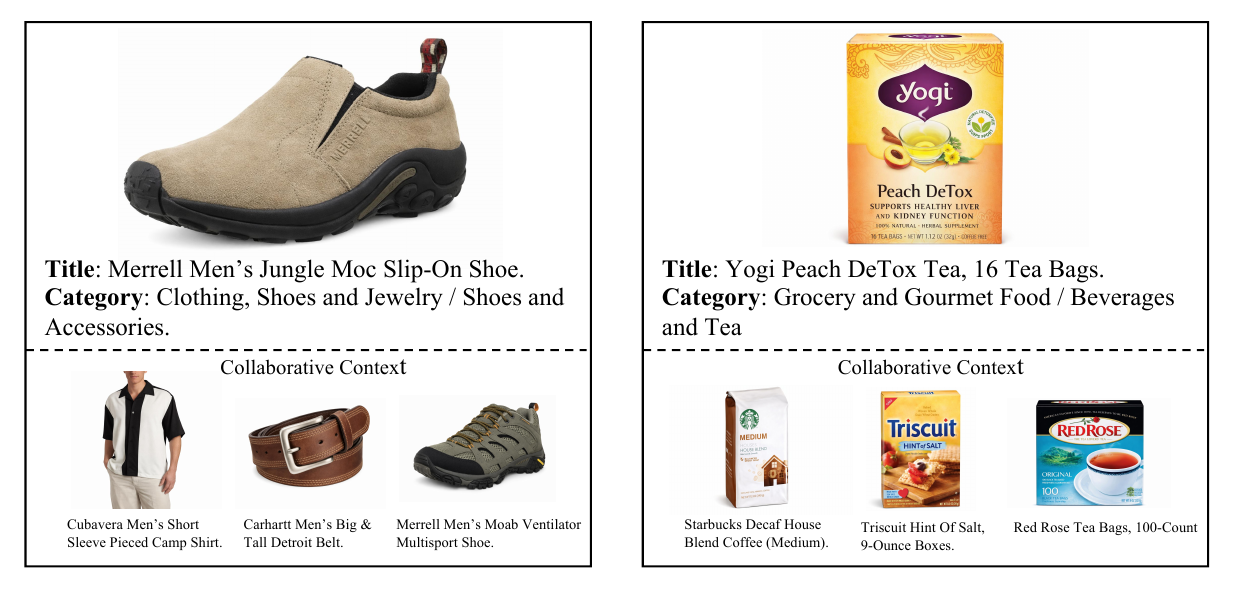}
  \end{minipage}\hfill
  \caption{Examples of visual semantic units on the Clothing and Food datasets.}
  \label{fig:case_study}
\end{figure}

By unifying textual, visual, and collaborative information into a card-style image representation $\mathcal{G}_i$, this design provides a rich and structurally consistent input to facilitate subsequent encoding and quantization. Representative examples of the constructed visual semantic units are presented in Figure \ref{fig:case_study}.

\subsubsection{\textbf{Unified Encoding with a Vision–Language Encoder.}}
Upon constructing the visual semantic unit $\mathcal{G}_i$, we treat it as a unified visual input to a pretrained vision--language encoder, denoted as $f(\cdot)$. This encoding process produces an $m$-dimensional semantic embedding:
\begin{equation}
\mathbf{z} = f(\mathcal{G}_i) \in \mathbb{R}^m .
\end{equation}

In our implementation, we adopt SigLIP2 \cite{SigLIP-2} as the vision--language encoder, which has been widely validated as a strong vision--language representation model. Its powerful holistic representation ability enables effective encoding of complex visual inputs, providing a stable and high-quality semantic foundation for subsequent modeling.

Although the unified embedding $\mathbf{z}_i$ effectively integrates multimodal semantic information and collaborative signals, its distribution in real-world recommendation scenarios is often highly non-uniform. This characteristic fundamentally limits the effectiveness of conventional uniform quantization strategies. To address this, we propose NU-RQ-VAE, which explicitly models such non-uniform semantic distributions.

\subsection{Non-Uniform Quantization Module}
The quantization module transforms embeddings $\mathbf{z}_i$ into discrete SID sequences, whose quality is crucial for downstream generative models to capture fine-grained item semantics. RQ-VAE~\cite{RQ-VAE} optimizes a global reconstruction loss (e.g., MSE), implicitly allocating codebook capacity uniformly across the latent space. This assumption often breaks in recommendation, where item semantics are highly non-uniform. Ignoring this non-uniformity leads to poor discriminability in high-density regions and inefficient codebook usage in low-density regions, degrading SID quality and generation stability. To address this issue, we propose a Non-Uniform Residual Quantized Variational Autoencoder (NU-RQ-VAE). We first review the standard RQ-VAE quantization procedure.

\subsubsection{\textbf{RQ-VAE Quantization Process}}
The RQ-VAE consists of an $Encoder(\cdot)$, a $Decoder(\cdot)$, and $K$ codebooks $\{\mathcal{C}_1,\dots,\mathcal{C}_K\}$. Specifically, for each code level $k\in\{1,\dots,K\}$, we have a codebook $\mathcal{C}_k=\{\bm{e}_i\}_{i=1}^N$, where $\bm{e}_i\in\mathbb{R}^{d}$ is a learnable code embedding and $N$ denotes the codebook size. 

For an item, we first project its embedding $\bm{z}$
into the latent space via the encoder $\bm{h} = Encoder(\bm{z})$. This $\bm{h}$ is then quantized through $K$ successive codebook levels via residual quantization. The residual quantization process can then be formally expressed as:
\begin{equation}
\left\{
\begin{aligned}
    &c_k=\mathop{\arg\min}_{i}\|\bm{r}_{k-1}-\bm{e}_i\|^2, \quad \bm{e}_i\in\mathcal{C}_k, \\
    &\bm{r}_k=\bm{r}_{k-1}-\bm{e}_{c_k},
\end{aligned}
\right.
\end{equation}
where $c_k$ is the assigned code index from the $k$-th level codebook, $\bm{r}_{k-1}$ is the semantic residual from the last level, and we set $\bm{r}_0=\bm{h}$. Intuitively, at each code level, RQ-VAE finds the most similar code embedding with the semantic residual and assigns the item with the corresponding code index. 
After the recursive quantization, we eventually obtain the SID $\Tilde{i}=[c_{1}, c_{2}, \dots, c_{K}]$, and the quantized embedding $\hat{\bm{h}}=\sum_{k=1}^{k}\bm{e}_{c_k}$. 
The quantized embedding $\hat{\bm{h}}$ is then decoded to the reconstructed semantic embedding $\hat{\bm{z}} = Decoder(\hat{\bm{h}})$.

The overall RQ-VAE loss consists of the reconstruction loss $\mathcal{L}_{\text{Recon}}$ and the quantization loss $\mathcal{L}_{\text{RQ}}$:
\begin{equation}
\left\{
\begin{aligned}
    &\mathcal{L}_{\text{Recon}} = \|\bm{z} - \hat{\bm{z}}\|^{2}, \\
    &\mathcal{L}_{\text{RQ}} =
    \sum_{k=1}^{K}
    \left(
    \|\text{sg}[\bm{r}_{k-1}] - \bm{e}_{c_k}\|^2
    + \mu \|\bm{r}_{k-1} - \text{sg}[\bm{e}_{c_k}]\|^2
    \right), \\
    &\mathcal{L}_{\text{Sem}} = \mathcal{L}_{\text{Recon}} + \mathcal{L}_{\text{RQ}},
\end{aligned}
\right.
\end{equation}
where $\text{sg}[\cdot]$ denotes the stop-gradient operation~\cite{VQ-VAE}, and $\mu$ controls the trade-off between optimizing the code embeddings and the encoder. The reconstruction loss $\mathcal{L}_{\text{Recon}}$ preserves essential semantic information by enforcing accurate reconstruction from the quantized embedding, while $\mathcal{L}_{\text{RQ}}$ minimizes residual errors across all quantization levels and jointly trains the encoder and codebooks~\cite{VQ-VAE}.

\subsubsection{\textbf{Non-Uniform RQ-VAE}}
\begin{figure}[tbp]
  \centering
    \centering
    \includegraphics[width=\linewidth]{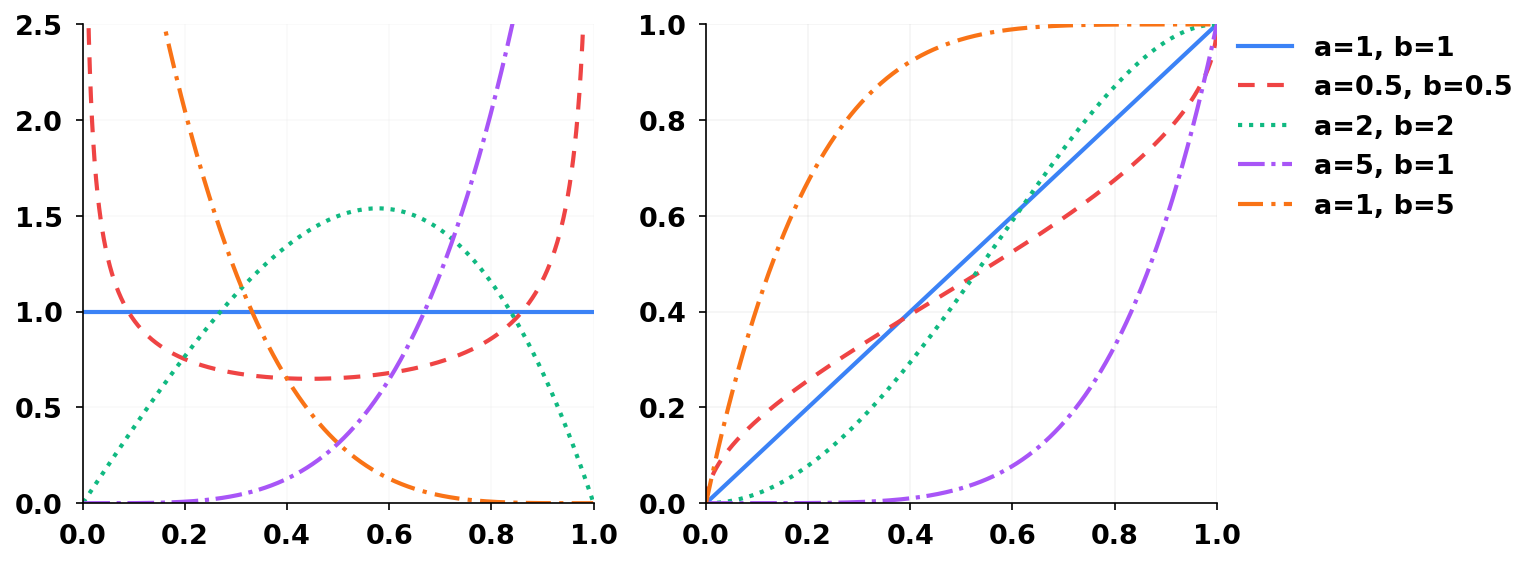}
  \caption{PDF (left) and CDF (right) of the Kumaraswamy distribution. The monotonic and invertible CDF enables reparameterizing non-uniform distributions into an approximately uniform space, making it suitable for distribution-aware quantization.}
  \label{fig:kumaraswamy}
\end{figure}
Inspired by non-uniform quantization \cite{nu1,nu2,nu3}, we introduce a learnable and invertible non-uniform transformation before residual quantization to reparameterize latent semantic representations. Quantization is then performed in an intermediate space where the transformed representations are approximately more uniform, making them more amenable to quantization; during decoding, the inverse transformation maps the quantized representations back to the original semantic space. Specifically, we introduce the following two invertible non-uniform transformation functions:

\noindent\textbf{The Kumaraswamy Distribution.}
To realize the proposed learnable and invertible non-uniform transformation, we construct a monotonic mapping that admits an efficient (closed-form) inverse, enabling efficient forward and inverse transforms. A natural choice is to parameterize the mapping via a distribution’s CDF and its quantile function. The Beta distribution is widely adopted due to its strong expressive power in modeling diverse data distributions over the interval $[0,1]$. However, its normalization term is defined by the Beta function, which lacks a closed-form expression, making it computationally expensive in practice. The Kumaraswamy distribution exhibits expressive power comparable to that of the Beta distribution, as illustrated in Figure~\ref{fig:kumaraswamy}. It is significantly more computationally efficient because its probability density function, cumulative distribution function, and quantile function all have closed-form expressions \cite{kumaraswamy_generalized_1980,jones_kumaraswamys_2009}. We adopt the Kumaraswamy distribution as the non-uniform transformation function. Its cumulative distribution function and quantile function are defined as follows:
\begin{align}
F_{\textsc{KS}}(x; a,b) &= 1 - \left(1 - x^{a}\right)^{b},
    \label{eq:kumaraswamy_cdf}
    \\
F_{\textsc{KS}}^{-1}(y; a,b) &= \left(1 - \left(1 - y\right)^{1/b}\right)^{1/a},
    \label{eq:kumaraswamy_icdf}
\end{align}
where $x \in [0,1]$ and $a,b \in \mathbb{R}{+}$. Although the Kumaraswamy distribution is conventionally defined over the open interval $(0,1)$, its cumulative distribution function remains well defined at the boundary points. Therefore, we safely extend its domain to the closed interval $[0,1]$. We normalize the values $x_i$ in $x$ as $(x_i - x_{\text{min}}) / (x_{\text{max}} - x_{\text{min}})$, where $x_{\text{min}} = \min_i x_i$ and $x_{\text{max}} = \max_i x_i$.

\begin{figure}[tbp]
  \centering
    \centering
    \includegraphics[width=\linewidth]{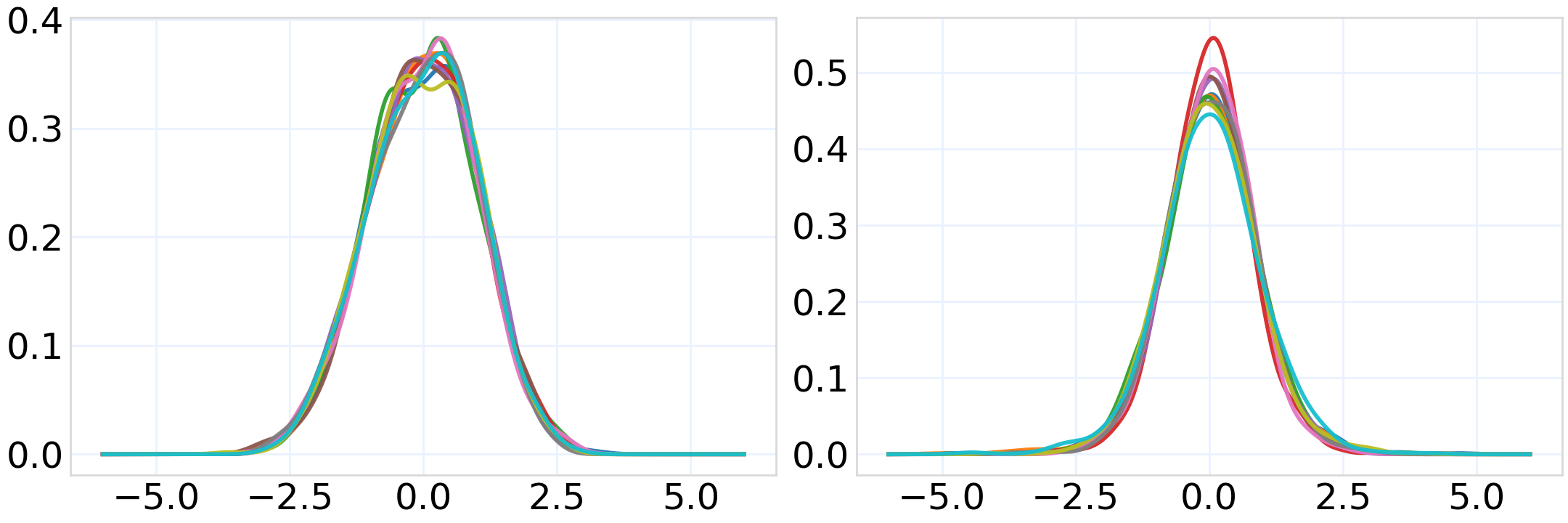}
  \caption{The left shows the embedding distributions of 10 samples from the T5 encoder on the Food dataset, and the right shows those from the SigLIP2 encoder. In both cases, the distributions exhibit an approximately bell-shaped form.}
  \label{fig:zhongxing}
\end{figure}

\noindent\textbf{The Scaled Logistic and Scaled Logit.}
To further understand the characteristics of semantic embeddings in recommendation scenarios, we analyze the distributions of item embeddings produced by different encoding methods. As shown in Figure \ref{fig:zhongxing}, regardless of the specific encoding approach, the semantic embeddings of individual items consistently exhibit an approximately bell-shaped distribution across dimensions.
Based on this observation, we further introduce scaled logistic and scaled logit functions as invertible transformations to perform non-uniform mapping on each semantic embedding. 
These functions naturally align with bell-shaped distributions while preserving monotonicity and invertibility. As a result, the quantization process becomes better adapted to the internal structure of the embedding space, enhancing its ability to discriminate densely populated semantic regions. 
The standard logistic and logit functions on $\mathbb{R} \to (0, 1)$ and $(0, 1) \to \mathbb{R}$, respectively, are defined as:
\begin{align}
    \operatorname{logistic} (x; \alpha, x_0) &=\left( 1 + \exp(-\alpha (x - x_0)) \right)^{-1} ,
    \label{eq:logistic}
    \\
    \operatorname{logit} (y; \alpha, x_0) &= \alpha^{-1} \log \left( \frac{y}{1 - y} \right) + x_0 .
    \label{eq:logit}
\end{align}

Let $x_{\text{min}} = \min_i x_i$ and $x_{\text{max}} = \max_i x_i$. We are interested in quantizing values in $[x_{\text{min}}, x_{\text{max}}]$. For this, we define scaled versions of these functions, i.e., $\operatorname{logistic}_{\textsc{scaled}} : [x_{\text{min}}, x_{\text{max}}] \to [0, 1]$ and $\operatorname{logit}_{\textsc{scaled}} : [0, 1] \to [x_{\text{min}}, x_{\text{max}}]$, as follows:
\begin{equation}
\begin{array}{l}
\displaystyle
\operatorname{logistic}_{\textsc{scaled}}(x;\alpha,x_0)
= \\[3pt]
\displaystyle
\frac{
    \operatorname{logistic}(\delta^{-1}x;\alpha,x_0)
    -
    \operatorname{logistic}(\delta^{-1}x_{\min};\alpha,x_0)
}{\Delta},
\end{array}
\end{equation}
\begin{equation}
\operatorname{logit}_{\textsc{scaled}}(y; \alpha, x_0)
=
\delta\,
\operatorname{logit}\!\left(
    \Delta y + \operatorname{logistic}(\delta^{-1} x_{\text{min}}; \alpha, x_0)
\right),
\label{eq:logit_scaled}
\end{equation}
where $\Delta = \operatorname{logistic}(x_{\text{max}}; \alpha, x_0) - \operatorname{logistic}(x_{\text{min}}; \alpha, x_0)$ and $\delta = x_{\text{max}}-x_{\text{min}}$. 
This scaling ensures that the logistic and logit transformations are well-defined, monotonic, and bounded on the prescribed domain, while preserving a consistent interpretation of the parameters across different input ranges. In particular, normalizing the input values $x$ by $\delta^{-1}$ makes the parameters $\alpha$ and $x_0$ invariant to the specific domain $[x_{\text{min}}, x_{\text{max}}]$, so that identical parameter values induce comparable nonlinear behavior across vectors. Equivalently, this normalization can be interpreted as a reparameterization with $\widetilde{\alpha} = \delta^{-1} \alpha$ and $\widetilde{x}_0 = \delta x_0$.

Based on the two non-uniform transformations introduced above, we define an invertible transformation pair $\big(\mathcal{T}(\cdot), \mathcal{T}^{-1}(\cdot)\big)$ and incorporate it into the RQ-VAE framework to better adapt residual quantization to the distributional characteristics of item representations in recommendation scenarios. Specifically, $\mathcal{T}(\cdot)$ and its inverse can be instantiated in one of the following two forms:
\begin{equation}
\mathcal{T}(\cdot),\; \mathcal{T}^{-1}(\cdot)
=
\begin{cases}
\big(F_{\textsc{KS}}(\cdot; a,b),\; F_{\textsc{KS}}^{-1}(\cdot; a,b)\big), \\
\big(\operatorname{logistic}_{\text{scaled}}(\cdot;\alpha,x_0),\;
\operatorname{logit}_{\text{scaled}}(\cdot;\alpha,x_0)\big).
\end{cases}
\end{equation}

Given a continuous embedding $\bm{h}$ produced by the RQ-VAE $Encoder(\cdot)$, we first map it into an approximately uniform intermediate space via $\bm{d}=\mathcal{T}(\bm{h})$, on which residual quantization is performed as described in Eq.~(3), yielding the quantized representation $\hat{\bm{d}}$. After quantization, we apply the inverse transformation prior to decoding, yielding $\hat{\bm{h}}=\mathcal{T}^{-1}(\hat{\bm{d}})$, so that the $Decoder(\cdot)$ operates in the original semantic space for representation reconstruction. This design enhances quantization by mapping non-uniform item representations into an approximately uniform space, while maintaining semantic consistency with the original representations through the inverse transformation.

To further encourage the consistency and stability of the non-uniform transformation, we introduce an additional NUQ (Non-uniform Quantization) consistency loss. Specifically, given the transformed representation $\bm{d}=\mathcal{T}(\bm{h})$, we apply the inverse transformation to reconstruct the original embedding $\tilde{\bm{h}} = \mathcal{T}^{-1}(\bm{d})$, and penalize the discrepancy between $\tilde{\bm{h}}$ and the original embedding $\bm{h}$. The NUQ consistency loss is defined as:
\begin{equation}
\mathcal{L}_{\text{NUQ}}
=
\left\| \mathcal{T}^{-1}(\mathcal{T}(\bm{h})) - \bm{h} \right\|_2^2.
\end{equation}

The overall quantization objective is formulated as a weighted combination of the residual quantization loss in Eq.~(4) and the NUQ consistency loss:
\begin{equation}
\mathcal{L}_{\text{overall}}
=\mathcal{L}_{\text{Sem}} + \lambda_{\text{NUQ}} \, \mathcal{L}_{\text{NUQ}},
\end{equation}
where $\lambda_{\text{NUQ}}$ is a hyper-parameter that controls the strength of the NUQ consistency regularization.

\subsection{Training and Recommendation}
\noindent\textbf{Training.}
The training of generative recommender models consists of two phases: quantization and model optimization. Each item is quantized into an SID $\tilde{\bm{i}} = [c_1, c_2, \ldots, c_K]$ using the well-trained NU-RQ-VAE. Based on these SIDs, user interaction sequences are translated into code sequences. We construct a training dataset
$\mathcal{D} = \{(x^{(\xi)}, y^{(\xi)})\}_{\xi=1}^{|\mathcal{D}|}$, where $x^{(\xi)} = [\tilde{\bm{i}}_1, \ldots, \tilde{\bm{i}}_M]$ denotes the historical interaction sequence, and $y^{(\xi)} = \tilde{\bm{i}}_{M+1}$ is the SID of the next interacted item. The model is trained in an autoregressive manner by minimizing the token-level negative log-likelihood:
\begin{equation}
\mathcal{L}
=
-\sum_{t=1}^{K}
\log P_{\theta}\!\left(
y_t^{(\xi)} \mid y_{<t}^{(\xi)},\, x^{(\xi)}
\right),
\end{equation}

\noindent\textbf{Recommendation.}
During inference, the generative recommender produces the code sequence of the next item in an autoregressive fashion. At each decoding step $t$, the model estimates a conditional probability distribution $P_{\theta}(v \mid y_{<t}^{(\xi)}, x)$ over the code vocabulary $\mathcal{V}$. Sequence decoding is carried out using beam search, which maintains the top-$B$ most probable partial sequences at each step to enhance recommendation quality.

\section{EXPERIMENTS}
\subsection{Experimental Setup}

\subsubsection{Datasets.} We conduct experiments on three categories Amazon \cite{amazon-2014} datasets: ``Grocery and Gourmet Food'' (Food), ``Cell Phones and Accessories'' (Phones), and ``Clothing, Shoes and Jewelry'' (Clothing). Following prior work \cite{SASRec,TIGER,LETTER}, we apply the 5-core filtering strategy to remove users and items with fewer than five interactions. To standardize the training process, each user’s interaction history is truncated or padded to a fixed length of 20 by retaining the most recent interactions. For data splitting, we adopt the widely used leave-one-out strategy \cite{LETTER,TIGER,SASRec}, where the most recent interaction of each user is used for testing, the second most recent for validation, and the remaining interactions are used for training. The detailed statistics of these datasets are reported in Table \ref{tab:dataStatistics}.

\begin{table}[t]
\small
\centering
\caption{
Dataset statistics. “Avg. len” denotes the mean length of user interaction sequences.
}
\label{tab:dataStatistics}  
 \resizebox{1.0\columnwidth}{!}{
\begin{tabular}{cccccc}
\toprule
Dataset & \#Users & \#Items  & \#Interaction &Sparsity &Avg.\emph{len} \\
\midrule
Food &14,681 &8,713 &151254 &99.882\% &10.30 \\
Phones &27,879 &10,429 &194,439 &99.933\% &6.97 \\
Clothing &39,387 &23,033 &278,677 &99.969\% &7.08 \\
\bottomrule
\end{tabular}
}
\end{table}

\subsubsection{Evaluation Metrics.}
We adopt Top-$K$ Recall (Recall@$K$) and Normalized Discounted Cumulative Gain (NDCG@$K$) as evaluation metrics, and assess recommendation performance with $K = 5, 10, 20$. We conduct a full-ranking \cite{TIGER,LETTER} evaluation over the entire item set.

\begin{table*}[t!]
\centering
\caption{Top-\textit{K} recommendation performance on three datasets. Best results are in bold, and best baselines are underlined. $^{*}$ indicates $p\leq0.01$ in paired t-tests against the best baseline. CARD$_K$ and CARD$_S$ denote variants with Kumaraswamy-based and scaled logistic/logit-based non-uniform quantization, respectively.}
\label{tab:exp}
\resizebox{\textwidth}{!}{%
\begin{tabular}{ccccccccccccc}
    \toprule
    \multicolumn{2}{c}{Setting} & \multicolumn{9}{c}{Baseline Methods} & \multicolumn{2}{c}{Ours}\cr 
    \cmidrule(lr){1-2} \cmidrule(lr){3-11} \cmidrule(lr){12-13}
    Dataset & \multicolumn{1}{c}{Metric} & GRU4Rec & HGN & BERT4Rec & SASRec & VQ-Rec & TIGER & LETTER & MQL4GRec & MACRec & CARD$_K$ & CARD$_S$ \cr
    \midrule
    \multirow{6}{*}{\rotatebox{90}{Food}} 
    & Recall@5 & 0.0345 & 0.0364 & 0.0325 & 0.0386 & 0.0423 & 0.0394 & 0.0437 & 0.0445 & \underline{0.0492} & 0.0520$^{*}$ & \textbf{0.0547}$^{*}$ \cr
    & Recall@10 & 0.0540 & 0.0575 & 0.0523 & 0.0596 & 0.0646 & 0.0617 & 0.0683 & 0.0697 & \underline{0.0779} & 0.0819$^{*}$  & \textbf{0.0853}$^{*}$ \cr
    & Recall@20 & 0.0849 & 0.0829 & 0.0806 & 0.0874 & 0.0982 & 0.0891 & 0.1004 & 0.1023 & \underline{0.1137} & 0.1212$^{*}$  & \textbf{0.1238}$^{*}$ \cr
    \cmidrule(lr){2-2}\cmidrule(lr){3-11}
    & NDCG@5 & 0.0218 & 0.0237 & 0.0209 & 0.0239 & 0.0275 & 0.0256 & 0.0283 & 0.0289 & \underline{0.0322} & 0.0333$^{*}$  & \textbf{0.0364}$^{*}$ \cr
    & NDCG@10 & 0.0280 & 0.0302 & 0.0264 & 0.0286 & 0.0351 & 0.0327 & 0.0362 & 0.0369 & \underline{0.0410} & 0.0428$^{*}$  & \textbf{0.0462}$^{*}$ \cr
    & NDCG@20 & 0.0358 & 0.0367 & 0.0340 & 0.0361 & 0.0426 & 0.0397 & 0.0439 & 0.0448 & \underline{0.0486} & 0.0527$^{*}$  & \textbf{0.0559}$^{*}$ \cr
    \midrule
    \multirow{6}{*}{\rotatebox{90}{Phones}} 
    & Recall@5 & 0.0483 & 0.0438 & 0.0450 & 0.0504  & 0.0511 & 0.0526 & 0.0544 & 0.0540 & \underline{0.0561} & \textbf{0.0596}$^{*}$  & 0.0585$^{*}$ \cr
    & Recall@10 & 0.0783 & 0.0671 & 0.0707 & 0.0762 & 0.0766 & 0.0794 & 0.0830 & 0.0838 & \underline{0.0850} & \textbf{0.0904}$^{*}$  & 0.0880$^{*}$ \cr
    & Recall@20 & 0.1102 & 0.1008 & 0.1034 & 0.1075 & 0.1095 & 0.1132 & 0.1179 & 0.1158 & \underline{0.1172} & \textbf{0.1284}$^{*}$  & 0.1250$^{*}$ \cr
    \cmidrule(lr){2-2}\cmidrule(lr){3-11}
    & NDCG@5 & 0.0310 & 0.0255 & 0.0276 & 0.0315 & 0.0318 & 0.0338 & 0.0353 & 0.0350 & \underline{0.0355} & 0.0385$^{*}$  & \textbf{0.0388}$^{*}$ \cr
    & NDCG@10 & 0.0407 & 0.0347 & 0.0365 & 0.0405 & 0.0417 & 0.0425 & 0.0443 & 0.0441 & \underline{0.0453} & \textbf{0.0486}$^{*}$  & 0.0474$^{*}$ \cr
    & NDCG@20 & 0.0519 & 0.0451 & 0.0464 & 0.0505 & 0.0492 & 0.0510 & 0.0536 & 0.0533 & \underline{0.0546} & \textbf{0.0578}$^{*}$ & 0.0570$^{*}$ \cr
    \midrule
    \multirow{6}{*}{\rotatebox{90}{Clothing}} 
    & Recall@5 & 0.0094 & 0.0140 & 0.0134 & 0.0141 & 0.0145 & 0.0152 & 0.0158 & 0.0170 & \underline{0.0175} &  \textbf{0.0191}$^{*}$ & 0.0187$^{*}$ \cr
    & Recall@10 & 0.0159 & 0.0213 & 0.0215 & 0.0223 & 0.0232 & 0.0257 & 0.0263 & 0.0273  & \underline{0.0281} & 0.0297$^{*}$ & \textbf{0.0301}$^{*}$\cr
    & Recall@20 & 0.0255 & 0.0316 & 0.0322 & 0.0330 & 0.0382 & 0.0418 & 0.0436 & 0.0441  & \underline{0.0450} & \textbf{0.0483}$^{*}$ & 0.0480$^{*}$\cr
    \cmidrule(lr){2-2}\cmidrule(lr){3-11}
    & NDCG@5 & 0.0061 & 0.0072 & 0.0074 & 0.0076 & 0.0081 & 0.0093 & 0.0092 & \underline{0.0097}  & 0.0096 & \textbf{0.0114}$^{*}$ & 0.0112$^{*}$ \cr
    & NDCG@10 & 0.0082 & 0.0097 & 0.0106 & 0.0102 & 0.0114 & 0.0113 & 0.0110 & \underline{0.0135}  & 0.0133 & \textbf{0.0150}$^{*}$ & 0.0145$^{*}$\cr
    & NDCG@20 & 0.0106 & 0.0122 & 0.0127 & 0.0130 & 0.0144 & 0.0167 & 0.0152 & 0.0167 & \underline{0.0170}  & \textbf{0.0192}$^{*}$ & 0.0185$^{*}$ \cr
    \bottomrule
\end{tabular}%
}
\end{table*}

\subsubsection{Implementation Details}
The encoder and decoder of the proposed NU-RQ-VAE are implemented as $3$-layer MLPs. For the codebook configuration, we follow the same setting as the baseline methods \cite{TIGER,LETTER}, using $K=4$ codebooks, each containing $N=256$ code vectors with a dimensionality of 32. The model is optimized using the AdamW optimizer~\cite{adam} with a learning rate of 0.001 and a batch size of 1024, and the weight hyperparameter $\lambda_{\text{NUQ}}$ is tuned within the range of $[0.1, 1]$. We implement the Transformer encoder–decoder architecture based on the T5 model~\cite{T5}. Both the Transformer encoder and decoder consist of 4 layers, with each layer including 6 self-attention heads of dimension 64. The T5 model is also trained using the AdamW optimizer \cite{adam}. 
For a fair comparison, MQL4GRec did not use pre-training on large-scale additional-category data.

Each visual semantic unit is constructed as a $512 \times 512$ image. The upper region is used to present the main item image, which is resized to fit the designated display area while preserving its aspect ratio and centered horizontally. The middle text region is used to display the item title, with optional category information when needed. The bottom region adopts a multi-column layout to present several collaborative neighbors, each consisting of a thumbnail and its corresponding title text. The specific layout and size configuration can be adjusted according to the content characteristics of different datasets.

\subsubsection{Baselines.} 
To evaluate the performance of our proposed model CARD, we compare it with the following baseline methods, including both traditional sequential recommendation methods and generative recommendation methods:

    \begin{itemize}[leftmargin=*, labelsep=2pt, align=left]
        \item \textbf{GRU4Rec} \cite{GRU4Rec} is a GRU-based sequential method that predicts the next item from user behavior sequences.
        \item \textbf{HGN} \cite{HGN} models users’ sequential interests by selecting key items and features through hierarchical gating.
        \item \textbf{BERT4Rec} \cite{BERT4Rec} leverages a bidirectional Transformer to model user interaction sequences by capturing contextual information from both directions.
        \item \textbf{SASRec} \cite{SASRec} builds on self-attention to model long-range dependencies in user interaction sequences. 
        \item \textbf{VQ-Rec} \cite{VQ-Rec} quantizes each item into SIDs via product quantization, and pools the SID embeddings to represent each item.
        \item \textbf{TIGER} \cite{TIGER} quantizes each item using RQ-VAE, then trains the model to autoregressively generate the next SID token by token, and uses beam search during inference.
        \item \textbf{LETTER} \cite{LETTER} is a learnable item tokenization method for generative recommendation that adaptively learns item identifiers by leveraging hierarchical semantics and collaborative signals.
        \item \textbf{MQL4GRec} \cite{MQL4GRec} leverages pre-training and fine-tuning to quantize multimodal information of items into SIDs, thereby constructing a unified representation across modalities and domains.
        \item \textbf{MACRec} \cite{MACRec} is a generative recommendation model that improves SID learning via cross-modal quantization and multi-aspect alignment. 
    \end{itemize}

\subsection{Overall Performance}
Table~\ref{tab:exp} reports the Top-$K$ recommendation performance of CARD and all baseline methods on three datasets. We observe that both variants of CARD (CARD$_K$ and CARD$_S$) consistently outperform existing baseline methods across all datasets and evaluation metrics. 
This demonstrates that CARD exhibits stable and robust performance advantages in recommendation scenarios. Compared with traditional sequential recommendation models (e.g., GRU4Rec, HGN, BERT4Rec, and SASRec), CARD achieves significant improvements on all evaluation metrics, indicating the clear benefits of formulating recommendation as SID-based generative sequence prediction. Moreover, CARD consistently outperforms existing generative recommendation methods across all datasets, including those relying on multimodal modeling such as MQL4GRec and MACRec. 

More specifically, the performance gains of CARD arise from two complementary designs. 
First, by unifying multimodal content and collaborative semantics at the encoding stage via visual semantic units, CARD mitigates the instability and semantic gap introduced by the “separate-then-fuse” paradigm under insufficient supervision, providing a coherent semantic basis for quantization. 
Second, the proposed non-uniform quantization explicitly accounts for the highly skewed semantic distributions in recommendation, enabling better codebook utilization, reducing generation bias, and improving quantization stability. Together, these two components allow heterogeneous information to be compressed into discrete SIDs in a more stable and expressive manner, leading to improvements in generative recommendation performance.

\subsection{Ablation Study}

\begin{table}[tbp]
\centering
\caption{Ablation study of \textbf{CARD$_K$} on the Food and Phones datasets.}
\begin{tabularx}{\linewidth}{lXXXX}
\toprule
\multirow{2}{*}{\textbf{Ablation}} 
& \multicolumn{2}{c}{\textbf{Food}} 
& \multicolumn{2}{c}{\textbf{Phones}} \\
\cmidrule(lr){2-3} \cmidrule(lr){4-5}
& R@5 & R@10 & R@5 & R@10 \\
\midrule
CARD$_K$        & 0.0520 & 0.0819 & 0.0596 & 0.0904 \\
w/o Unit-V     & 0.0452 & 0.0711 & 0.0564 & 0.0835 \\
w/o Unit-T     & 0.0463 & 0.0732 & 0.0557 & 0.0820 \\
w/o Unit-C     & 0.0491 & 0.0768 & 0.0571 & 0.0863 \\
w/o Unit       & 0.0432 & 0.0683 & 0.0542 & 0.0809 \\
w/o NUT        & 0.0478 & 0.0745 & 0.0560 & 0.0841 \\
\bottomrule
\end{tabularx}
\label{tab:ablation_cardk}
\end{table}

\begin{table}[tbp]
\centering
\caption{Ablation study of \textbf{CARD$_S$} on the Food and Phones datasets.}
\begin{tabularx}{\linewidth}{lXXXX}
\toprule
\multirow{2}{*}{\textbf{Ablation}} 
& \multicolumn{2}{c}{\textbf{Food}} 
& \multicolumn{2}{c}{\textbf{Phones}} \\
\cmidrule(lr){2-3} \cmidrule(lr){4-5}
& R@5 & R@10 & R@5 & R@10 \\
\midrule
CARD$_S$        & 0.0547 & 0.0853 & 0.0585 & 0.0880 \\
w/o Unit-V     & 0.0485 & 0.0771 & 0.0558 & 0.0792 \\
w/o Unit-T     & 0.0472 & 0.0754 & 0.0551 & 0.0793 \\
w/o Unit-C     & 0.0525 & 0.0821 & 0.0565 & 0.0848 \\
w/o Unit       & 0.0456 & 0.0702 & 0.0545 & 0.0819 \\
w/o NUT        & 0.0513 & 0.0795 & 0.0564 & 0.0831 \\
\bottomrule
\end{tabularx}
\label{tab:ablation_cards}
\end{table}

\begin{table*}[htbp]
\caption{
Performance of Non-Uniform Quantization with Different Quantizers.
}
\resizebox{1.99\columnwidth}{!}{
\begin{tabular}{l|cc|cc|cc}
\toprule
\multirow{2}{*}{Methods}
& \multicolumn{2}{c|}{Food}
& \multicolumn{2}{c|}{Phones}
& \multicolumn{2}{c}{Clothing} \\
\cmidrule(r){2-3}\cmidrule(r){4-5}\cmidrule(r){6-7}
& Recall@5 & Recall@10
& Recall@5 & Recall@10
& Recall@5 & Recall@10 \\
\midrule

R-VQ
& 0.0422 & 0.0638
& 0.0495 & 0.0773
& 0.0145 & 0.0245 \\

NU-R-VQ$_{K}$
& 0.0461 (+9.24\%) & 0.0729 (+14.26\%)
& 0.0539 (+8.89\%) & 0.0812 (+5.05\%)
& 0.0159 (+9.66\%) & 0.0268 (+9.39\%) \\

NU-R-VQ$_{S}$
& 0.0472 (+11.85\%) & 0.0722 (+13.17\%)
& 0.0546 (+10.30\%) & 0.0823 (+6.47\%)
& 0.0155 (+6.90\%)  & 0.0261 (+6.53\%) \\
\midrule

RQ-VAE
& 0.0394 & 0.0617
& 0.0526 & 0.0794
& 0.0152 & 0.0257 \\

NU-RQ-VAE$_{K}$
& 0.0442 (+12.18\%) & 0.0693 (+12.32\%)
& 0.0556 (+5.70\%)  & 0.0838 (+5.54\%)
& 0.0164 (+7.89\%)  & 0.0273 (+6.23\%) \\

NU-RQ-VAE$_{S}$
& 0.0456 (+15.74\%) & 0.0713 (+15.56\%)
& 0.0551 (+4.75\%)  & 0.0842 (+6.05\%)
& 0.0160 (+5.26\%)  & 0.0269 (+4.67\%) \\

\bottomrule
\end{tabular}
}
\label{tab:rq_model_type}
\end{table*}

\begin{table}[htbp]
\centering
\caption{Comparison between text-only encoding variants.}
\setlength{\tabcolsep}{10pt}  
\renewcommand{\arraystretch}{1.15} 
\begin{tabular}{lcccc}
\toprule
\multirow{2}{*}{Variant} 
& \multicolumn{2}{c}{Food}
& \multicolumn{2}{c}{Phones} \\
\cmidrule(lr){2-3} \cmidrule(lr){4-5}
& R@5 & R@10 & R@5 & R@10 \\
\midrule
TIGER+T5        & 0.0394  & 0.0617 & 0.0526 & 0.0794 \\
TIGER+Unit-T    & 0.0389 & 0.0625 & 0.0531 & 0.0788 \\
\bottomrule
\end{tabular}
\label{tab:text_variant}
\end{table}

To further validate the effectiveness of each component in CARD, we conduct ablation studies on the Food and Phones datasets for both CARD$_K$ and CARD$_S$, as reported in Tables~\ref{tab:ablation_cardk} and \ref{tab:ablation_cards}. Removing visual (w/o Unit-V), textual (w/o Unit-T), or collaborative signals (w/o Unit-C) consistently degrades performance, indicating that these information sources provide complementary semantics for stable and discriminative SID learning. Replacing the visual semantic unit with a text-only representation (w/o Unit) leads to the most severe performance drop, highlighting the benefit of unifying multimodal content and collaborative semantics at the encoding stage. We further examine the role of non-uniform quantization. Removing the non-uniform transformation and NUQ regularization loss (w/o NUT), which reduces the model to a standard RQ-VAE, consistently harms performance across all settings, suggesting that conventional quantization struggles with the highly skewed semantic distributions in recommendation scenarios. Overall, the visual semantic unit and non-uniform quantization act complementarily, jointly enabling robust and effective SID construction.

\begin{figure}[tbp]
  \centering
    \centering
    \includegraphics[width=\linewidth]{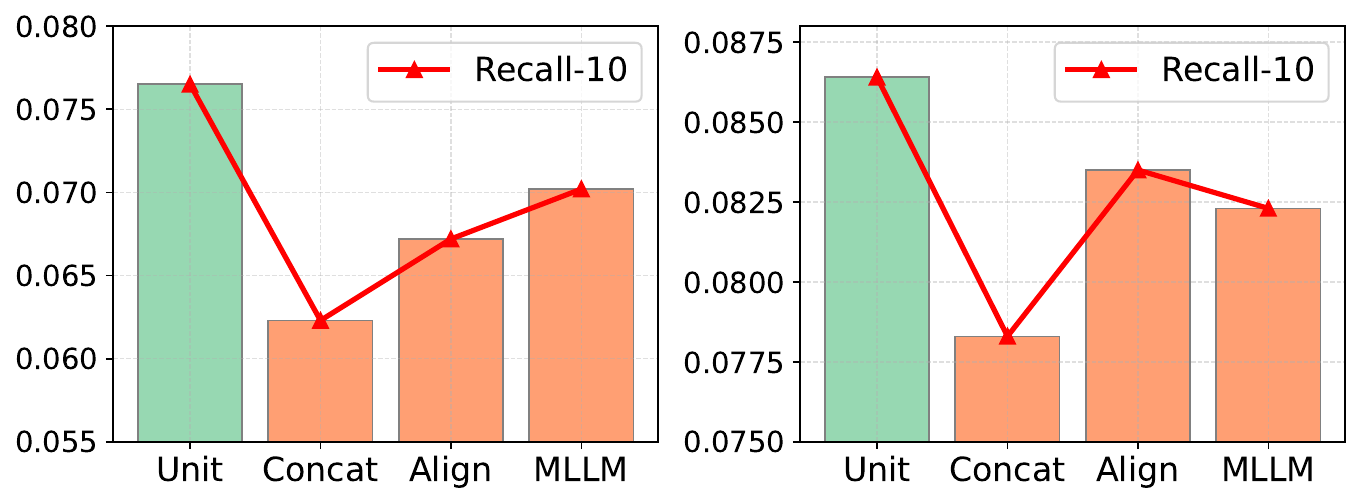}
  \caption{Comparison of different multimodal fusion strategies in terms of Recall@10. The left corresponds to the Food dataset, and the right corresponds to the Phones dataset.}
  \label{fig:fusion}
\end{figure}

\subsection{Further Analysis}

\subsubsection{Fusion Strategy Variants Analysis.}
To further examine the effectiveness and practicality of different multimodal fusion strategies for SID learning, we compare visual semantic unit (Unit) with three representative alternatives:
(1) Concatenation-based Fusion (Concat), which independently encodes
textual, visual, and collaborative signals and fuses them via feature concatenation;
(2) Contrastive-aligned Fusion (Align), which aligns heterogeneous
representations through contrastive pretraining before fusion; and
(3) MLLM-based Joint Encoding (MLLM), which uses a multimodal large
language model (LLaMA3-8B) to jointly encode textual and visual inputs, extracting hidden representations as item embeddings.

As shown in Figure \ref{fig:fusion}, Unit consistently achieves the best Recall@10 performance while maintaining a significantly simpler modeling pipeline. Concatenation-based fusion lacks unified semantic organization and is therefore prone to semantic gaps, whereas contrastive-aligned fusion may disrupt the original semantic structures of individual modalities during alignment, thereby weakening modality-specific information. MLLM representations are more geared toward high-level semantic abstraction for generative understanding. As a result, their joint encoding process often involves strong information compression, making it difficult to fully preserve fine-grained cues from both images and text. Unit unifies heterogeneous information into visual semantic units prior to encoding, which avoids semantic gaps while preserving modality-specific critical details. As a result, it produces representations with continuous semantic structure, stable distributions, and improved quantization-friendliness.

\subsubsection{Unit-based Text Encoding vs. Text-only Encoding.}

To examine whether rendering textual information into visual semantic units introduces semantic loss, we compare two text-only encoding variants under the TIGER framework. TIGER+T5 encodes item text using a T5 text encoder, while TIGER+Unit-T renders the same text into visual semantic units and encodes it with SigLIP2. As shown in Table~\ref{tab:text_variant}, the two variants achieve highly comparable performance on both the Food and Phones datasets, indicating that visual-semantic encoding does not lead to noticeable semantic degradation. This can be attributed to the following factors. First, SigLIP2, pretrained on large-scale image–text pairs, can reliably extract textual semantics from visual representations. Second, the visual semantic unit adopts a clean background, fixed layout, and structured fields, enabling stable and consistent rendering of textual content. Finally, item text in recommendation scenarios is typically short and semantically concise, allowing its core information to be preserved even when encoded in visual form. Minor differences between encoding methods are further mitigated during subsequent quantization.

\begin{figure}[tbp]
  \centering
  \begin{minipage}{.36\linewidth}
    \centering
    \includegraphics[width=\linewidth]{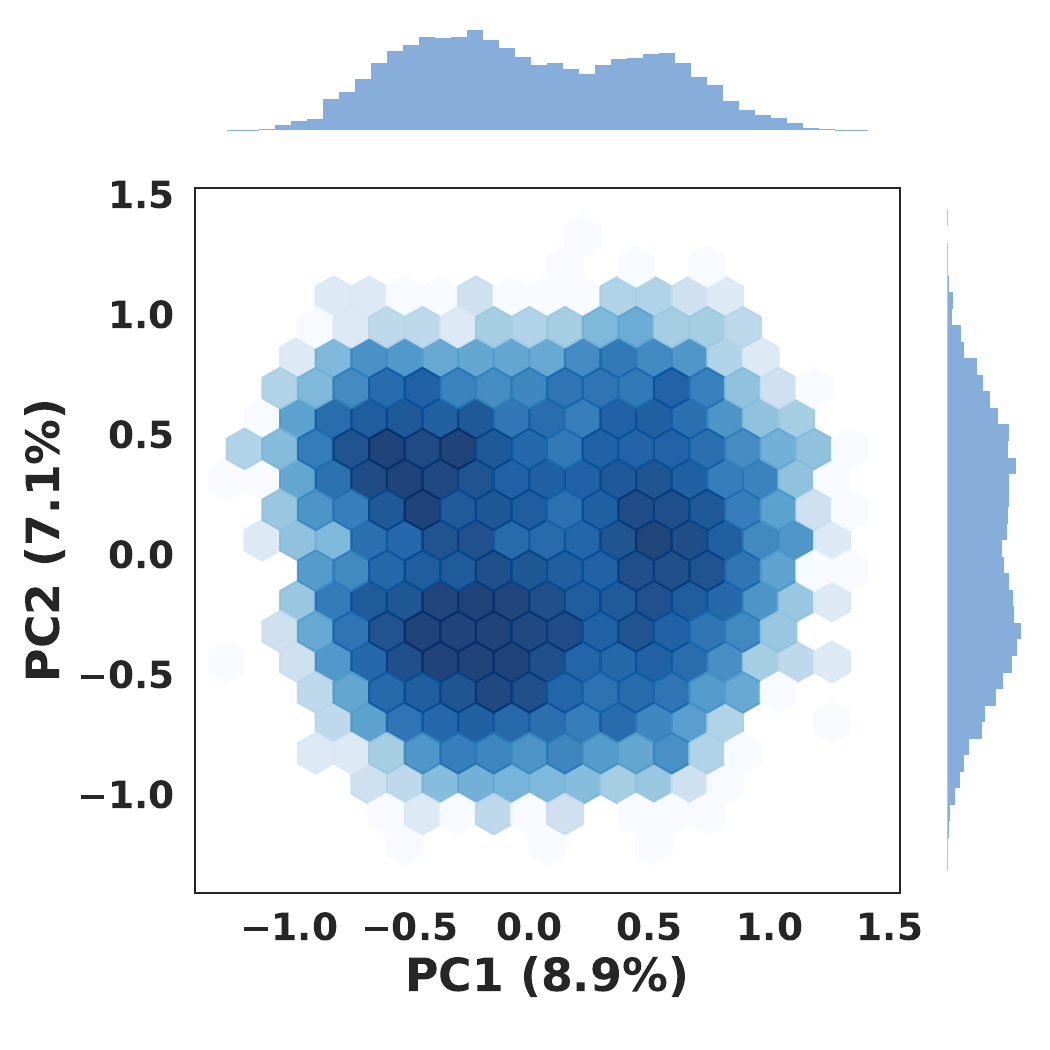}
  \end{minipage}\hfill
  \begin{minipage}{.64\linewidth}
    \centering
    \includegraphics[width=\linewidth]{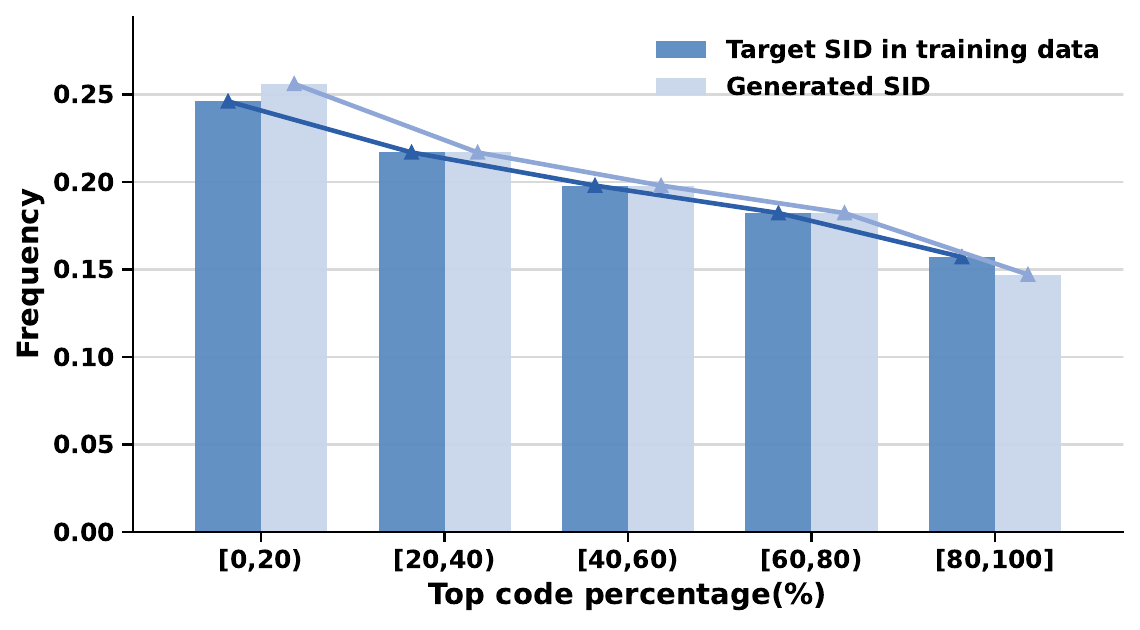}
  \end{minipage}
  \caption{The left shows item embeddings after non-uniform transformation, exhibiting a more compact and balanced semantic distribution. The right shows that the non-uniform transformation leads to more balanced codeword usage, reducing generation bias. }
  \label{fig:maben2}
\end{figure}

\begin{figure}[tbp]
  \centering
  \begin{minipage}{\linewidth}
    \centering
    \includegraphics[width=\linewidth]{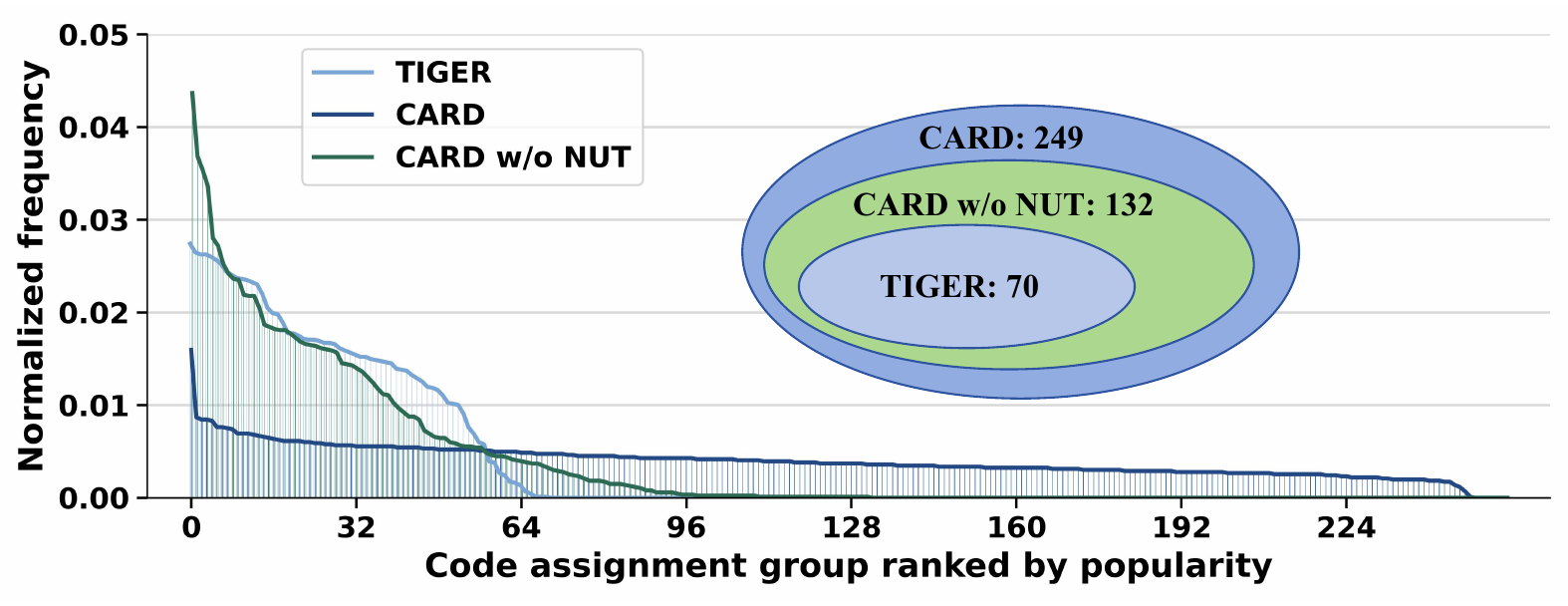}
  \end{minipage}
  \caption{Compared with TIGER and CARD w/o NUT, CARD with non-uniform transformation significantly increases the number of effectively utilized codewords, indicating more balanced codebook usage.}
  \label{fig:maben}
\end{figure}

\subsubsection{Plug-and-Play Property of Non-Uniform Quantization.}
To evaluate the plug-and-play property of non-uniform quantization, we apply the proposed transformation to different quantizers, including R-VQ and RQ-VAE. As shown in Table~\ref{tab:rq_model_type}, introducing non-uniform quantization consistently improves performance across all datasets and metrics. For both R-VQ and RQ-VAE, their non-uniform variants achieve stable and notable gains, with relative improvements of up to 15.74\%. These results indicate that the proposed non-uniform transformation is quantizer-agnostic and can be seamlessly integrated into different quantization frameworks, effectively correcting skewed semantic distributions and improving SID quality.

\subsubsection{Effect of Non-Uniform Quantization.}
As shown in Figure~\ref{fig:maben2}, after applying the proposed non-uniform transformation, the embeddings of visual semantic units form a more balanced distribution in the PCA space, alleviating dense clustering and long-tail sparsity that hinder effective quantization. Such a distribution is more favorable for subsequent discrete quantization.
Furthermore, as illustrated in Figure~\ref{fig:maben}, compared with TIGER and CARD w/o NUT, the full CARD model significantly increases the number of effectively utilized codewords. The distributions of generated SIDs better align with those of target SIDs, indicating more balanced codebook usage and reduced generation bias, which in turn contributes to improved recommendation performance.

\section{CONCLUSION AND DISCUSSION}
In this paper, we propose CARD, a novel generative recommendation framework designed to address two challenges in SID construction: insufficient supervision for heterogeneous information fusion and codeword imbalance caused by non-uniform semantic embeddings. CARD introduces a visual semantic unit that unifies textual, visual, and collaborative signals into a structured visual representation prior to encoding, enabling holistic semantic modeling without relying on explicit fusion mechanisms. This design provides a semantically coherent and quantization-friendly representation space for SID-based generative recommendation. To further mitigate the skewed semantic distributions commonly observed in real-world recommendation scenarios, we propose NU-RQ-VAE, which incorporates a learnable and invertible non-uniform transformation into residual quantization. Extensive experiments on multiple datasets demonstrate that both variants of CARD consistently outperform existing baseline methods, while the proposed non-uniform transformation module exhibits strong plug-and-play capability and can be generalized to different quantization methods. While CARD demonstrates strong effectiveness, the current visual semantic unit relies on a certain amount of human prior in representation structure design, and we plan to explore more automated construction in future work.

\begin{acks}
This research was supported by the National Natural Science Foundation of China (62402093), the Sichuan Science and Technology Program (2025ZNSFSC0479), and the Fundamental Research Funds for the Central Universities (JBK202511020). This research was also supported by the Center for HPC, University of Electronic Science and Technology of China.
\end{acks}

\bibliographystyle{ACM-Reference-Format}
\bibliography{sample-base}
\end{document}